\documentclass{ws-ijmpd}
\usepackage[super,compress]{cite}

\usepackage{graphicx, graphbox, epsfig, amssymb} 
\usepackage{amsmath, amsfonts}
\usepackage{bm} 
\usepackage{enumitem}
\usepackage[usenames]{color}
\definecolor{navyblue}{rgb}{0.0, 0.0, 0.5}
\usepackage[linktocpage,colorlinks=true,allcolors=navyblue]{hyperref}
\usepackage[caption=false]{subfig}
\usepackage[utf8]{inputenc}
\usepackage{tensor}
\usepackage{soul}

\def\beq{\begin{equation}}
\def\eeq{\end{equation}}

\def\bea{\begin{eqnarray}}
\def\eea{\end{eqnarray}}

\newcommand{\cF}{\mathcal{F}}
\newcommand{\cA}{\mathcal{A}}
\newcommand{\Ftil}{\widetilde{F}}
\newcommand{\bF}{\bm{\mathcal{F}}}
\newcommand{\mbar}{\overline{m}}

\newcommand{\barh}{\overline{h}}

\newcommand{\ff}{\mathfrak{f}}
\newcommand{\hh}{\mathfrak{h}}

\begin{document}

\title{\large Geometrical optics for scalar, electromagnetic and gravitational waves on curved spacetime}

\author{Sam R.~Dolan}

\address{Consortium for Fundamental Physics, \\
School of Mathematics and Statistics, University of Sheffield, \\
Hicks Building, Hounsfield Road, Sheffield S3 7RH, United Kingdom}

\maketitle

\begin{history}

\end{history}

\begin{abstract}
The geometrical-optics expansion reduces the problem of solving wave equations to one of solving transport equations along rays. Here we consider scalar, electromagnetic and gravitational waves propagating on a curved spacetime in general relativity. We show that each is governed by a wave equation with the same principal part. It follows that: each wave propagates at the speed of light along rays (null generators of hypersurfaces of constant phase); the square of the wave amplitude varies in inverse proportion to the cross section of the beam; and the polarization is parallel-propagated along the ray (the Skrotskii/Rytov effect). We show that the optical scalars for a beam, and various Newman-Penrose scalars describing a parallel-propagated null tetrad, can be found by solving transport equations in a second-order formulation. Unlike the Sachs equations, this formulation makes it straightforward to find such scalars beyond the first conjugate point of a congruence, where neighbouring rays cross, and the scalars diverge. We discuss differential precession across the beam which leads to a modified phase in the geometrical-optics expansion. 
\end{abstract}


\ccode{PACS numbers:}

\date{\today}

\maketitle

\section{Introduction\label{sec:intro}}
One key consequence of General Relativity is that light rays are deflected by gravitational fields. A ray passing close to a body of mass $M$ is, to first approximation, bent through the Einstein deflection angle $\Delta \phi_{E} = 4GM / c^2 b$, where $G$, $c$ and $b$ are the gravitational constant, the speed of light, and the impact parameter, respectively. Eddington's eclipse experiment of 1919, and its successors, established that the Einstein angle matches observational data -- rather than e.g.~the Soldner angle $\Delta \phi_{E}/2$ of Newtonian theory. 

A second key consequence of General Relativity is the existence of gravitational waves, which was confirmed in 2015 by the observation of characteristic ``chirps'' from binary black hole mergers \cite{TheLIGOScientific:2017qsa}. In Einstein's theory, gravitational waves propagate at exactly the speed of light, whereas in alternative theories this is not necessarily the case. The recent observation of the binary neutron-star merger in both electromagnetic waves and gravitational waves was the first such multi-messenger result. The gamma ray-burst arrived within $\sim 2$s of the gravitational wave peak, after propagating for $\sim 10^9$ years, confirming that the propagation speeds of electromagnetic and gravitational waves are the same (at least to $1$ part in $\sim 10^{16}$) \cite{Monitor:2017mdv}. 

Our knowledge of the cosmos is built up from inferences drawn from observations of waves that have travelled over cosmological distances across a dynamical curved spacetime. Astronomers do not typically analyse Maxwell's fundamental equations directly, though. To account for the gravitational lensing of light, for example, it suffices to employ a (leading-order) geometrical-optics approximation \cite{Kristian:1965sz, Seitz:1994xf, Perlick:2004tq}. In this approximation, the gradient of the phase is tangent to a light ray and, in vacuum, a light ray is a null geodesic of the spacetime. The wave's square amplitude varies in inverse proportion to the transverse area of the beam, so that flux is conserved in vacuum. The polarization of the wave is parallel-propagated along the ray, in a phenomenon known as gravitational Faraday rotation \cite{Ishihara:1987dv} or the Skrotskii effect \cite{Skrotskii:1957, Sereno:2005} (also known as the Rytov effect).

Geometrical optics is a widely-used approximation scheme based around one fundamental assumption: that the wavelength (and inverse frequency) is significantly shorter than all other characteristic length (and time) scales \cite{KlineKay, BornWolf:2013}, such as the spacetime curvature scale(s) \cite{Perlick:2004tq}. For the gravitational lensing of electromagnetic radiation, this is typically a safe assumption. Even for gravitational waves -- which are generated by bulk motions of compact objects \cite{Abbott:2016bqf} with long wavelengths (e.g.~$\lambda \sim 10^{7} \text{m}$ for GW150914) -- this assumption is still typically a good one. However, there are scenarios such as the scattering of waves by black holes \cite{Dolan:2008kf, Leite:2017zyb}, where the short-wavelength assumptions break down. Higher-order corrections to geometrical-optics can, in principle, be found in a systematic way \cite{Ehlers:1967, Anile:1976gq, Dolan:2018ydp}.

In this paper we review the theory of geometrical optics at leading-order, for scalar-field, electromagnetic and gravitational waves. Our quantities of interest are $\Phi$, the scalar field; $F_{ab}$, the Faraday tensor; and $h_{ab}$ and $C_{abcd}$, the metric perturbation and Weyl tensors, respectively.

This paper is organised as follows. In Sec.~\ref{sec:foundations} we review the free field equations (\ref{sec:field}), and then the scalar (\ref{sec:scalar}), electromagnetic (\ref{sec:EM}) and gravitational-wave (\ref{sec:GW}) cases in more detail. In Sec.~\ref{sec:go}, we introduce the geometrical optics expansion, again for the scalar (\ref{sec:go-scalar}), electromagnetic (\ref{sec:go-em}) and gravitational-wave (\ref{sec:go-gw}) cases. In Sec.~\ref{sec:lensing} we examine how geometrical-optics informs gravitational lensing theory. Neighbouring rays are subject to geodesic deviation (\ref{sec:geodesic-deviation}), and a beam is described by optical \& Newman-Penrose (NP) scalars (\ref{subsec:optical-scalars}). We develop a transport-equation approach for finding these scalars, and examine the effect of differential precession on the wave's phase (\ref{subsec:diffprecession}). We conclude in Sec.~\ref{sec:discussion} with a discussion.

\emph{Conventions:} Here $g_{ab}$ is a metric with signature $-+++$. Units are such that the gravitational constant $G$ and the speed of light $c$ are equal to $1$. Indices are lowered (raised) with the metric (inverse metric), i.e., $u_a = g_{ab} u^b$ ($u^a = g^{ab} u_b$). Einstein summation convention is assumed. The metric determinant is denoted $g = \text{det}\,g_{ab}$. The letters $a, b, c, \ldots$ are used to denote \emph{spacetime} indices running from $0$ (the temporal component) to $3$, whereas letters $i,j,k,\ldots$ denote \emph{spatial} indices running from $1$ to $3$. The Levi-Civita tensor is $\varepsilon_{abcd} \equiv \sqrt{-g} [a b c d]$, with $[a b c d]$ the fully anti-symmetric Levi-Civita symbol such that $[0 1 2 3] = +1$. The covariant derivative of $X_b$ is denoted by $\nabla_a X_b$ or equivalently $X_{b;a}$, and the partial derivative by $\partial_a X_b$ or $X_{b,a}$. The symmetrization (anti-symmetrization) of indices is indicated with round (square) brackets, e.g.~$X_{(ab)} = \frac{1}{2} (X_{ab} + X_{ba})$ and $X_{[ab]} = \frac{1}{2} (X_{ab} - X_{ba})$. $\{k^a, n^a, m^a, \mbar^a\}$ denote the legs of a (complex) null tetrad. Complex conjugation is denoted with an over-line, or alternatively, with an asterisk: $\mbar^a = m^{a\ast}$.

\section{Fundamentals\label{sec:foundations}}

\subsection{Field equations\label{sec:field}}
We take as our starting point the action
$
S = \int \left( \mathcal{L}_{\Phi} + \mathcal{L}_{F} + \mathcal{L}_{G} + \mathcal{L}_{M} \right) \sqrt{-g} d^4x ,
$
where the free-field Lagragians are $\mathcal{L}_{\Phi}  = -\frac{1}{2} g^{ab} \nabla_a \Phi^\ast \nabla_b \Phi$, $\mathcal{L}_{F}  = -\frac{1}{4} F_{ab} F^{ab}$, and $\mathcal{L}_{G} = R / 16 \pi G$. Here $\Phi$ is a scalar field, $F_{ab}$ is the Faraday tensor for the electromagnetic field, and $R = g^{ab} R_{ab} = g^{ab} g^{cd} R_{acbd}$ is the Ricci scalar, with $R_{ab}$ and $R_{abcd}$ the Ricci and Riemann tensors, respectively. The latter is constructed from the metric $g_{ab}$ and its derivatives up to second order. The metric determinant is $g$. The final term $\mathcal{L}_{M}$ couples the free fields to the matter sector.

The Euler-Lagrange equations yield three field equations, viz.
\begin{subequations}
\begin{align}
\Box \Phi &= 0 ,  \label{eq:kg} \\
\nabla_b F^{ab} &= J^a ,  \label{eq:Maxwell1} \\
G_{ab} \equiv R_{ab} - \frac{1}{2} g_{ab} R &= 8 \pi G \, T_{ab} ,
\end{align}
\end{subequations}
where $\Box \equiv g^{ab} \nabla_a \nabla_b$. 
Here $J^a$ and $T_{ab}$ are four-current and stress-energy, respectively, which are divergence-free by construction: $\nabla_a J^a = 0 = \nabla_a T^{ab}$. 

The Faraday tensor is antisymmetric in its indices, i.e., $F_{(ab)} \equiv \frac{1}{2} \left( F_{ab} + F_{ba} \right) = 0$. The Riemann tensor is antisymmetric in both pairs of indices $R_{abcd} = -R_{bacd} = R_{badc}$, and has the additional symmetries $R_{abcd} = R_{cdab}$ and $R_{a[bcd]} = \frac{1}{3} \left( R_{abcd} + R_{acdb} + R_{adbc} \right) = 0$. The Faraday tensor and Riemann tensor obey Bianchi identities, viz.
\begin{align}
\nabla_{[a} F_{bc]} = 0 , \quad \quad 
\nabla_{[a} R_{bc] de} = 0 .
\end{align}

\subsection{Electromagnetism and gravitation}
We now consider the fields in more detail.

\subsubsection{Scalar field\label{sec:scalar}}
The stress-energy of the scalar field is
\beq
T^{(\Phi)}_{ab} = \nabla_{(a} \Phi^\ast \nabla_{b)} \Phi - \frac{1}{2} g_{ab} \nabla_c \Phi^\ast \nabla^c \Phi  .  \label{eq:scalar-se}
\eeq

\subsubsection{Electromagnetic field\label{sec:EM}}
The electric and magnetic fields at a point in spacetime depend on the choice of Lorentz frame. An observer with (unit) tangent vector $u^a$ and (orthonormal) spatial frame $e^a_i$ `sees' an electric field $E_i = F_{a b} e_i^a u^b$ and a magnetic field $B_i = \Ftil_{a b} e_i^a u^b$. Here $\Ftil_{a b}$ is the Hodge dual \cite{Stephani:2003tm} of the Faraday tensor, defined by
\beq
\Ftil_{a b} \equiv \frac{1}{2} \varepsilon_{abcd} F^{cd} ,
\eeq
where $\varepsilon_{abcd}$ is the Levi-Civita tensor. (It follows that $\overset{\approx}{X}_{ab} = -X_{ab}$ for any two-form $X_{ab}$.) 

It is convenient to introduce a complexified version of the Faraday tensor,
\beq
\cF_{ab} \equiv F_{ab} + i \widetilde{F}_{ab}.
\eeq
The complex tensor $\cF_{ab}$ is \emph{self-dual}, by virtue of the property $\widetilde{\cF}_{ab} = -i \cF_{ab}$. From its definition, it follows that $\cF^\ast_{ab} \cF^{ab} = 0$, where ${}^{\ast}$ denotes complex conjugation.
We may also introduce a complex three-vector $\bF$ with components $\cF_i \equiv \cF_{ab} e_i^a u^b$, whose real and imaginary parts yield the (observer-dependent) electric and magnetic fields, 
$\bF = \mathbf{E} - i \mathbf{B}$.
%
%
The complex scalar quantity 
\beq
\Upsilon \equiv -\frac{1}{8} \cF_{ab} \cF^{ab} = \frac{1}{2} \bF \cdot \bF \label{eq:frame-inv}
\eeq 
is frame-invariant. Its real and imaginary parts yield the well-known frame-invariants $\frac{1}{2}(E^2 - B^2)$ and (minus) $\mathbf{E} \cdot \mathbf{B}$, respectively \cite{Dennison:2012vf}. A Faraday field with $\Upsilon = 0$ is called \emph{null}. In the null case, any observer sees electric and magnetic fields are orthogonal and of equal magnitude. 

The Bianchi identity $\nabla_{[a} F_{bc]} = 0$ is equivalent to $\nabla_b \Ftil^{ab} = 0$. In the language of forms, $F$ is closed ($dF = 0$ by the Bianchi identity), and thus by Poincar\'e's lemma, $F$ must be locally exact ($F = dA$). Thus, the Faraday tensor can be written in terms of a vector potential $A_a$ as
\beq
F_{ab} \equiv 2 \nabla_{[a} A_{b]} .
\eeq
Due to antisymmetry, it follows that $F_{ab} = 2 \partial_{[a} A_{b]} = \frac{\partial A_b}{\partial x^a} - \frac{\partial A_a}{\partial x^b}$. The Faraday tensor is invariant under gauge transformations of the form $A_{a} \rightarrow A^\prime_a = A_a + \partial_a \chi$, where $\chi$ is any scalar field.

By taking a derivative of the first equation of (\ref{eq:Maxwell1}), re-ordering covariant derivatives, and applying the Bianchi identity, one may obtain a wave equation in the form
\beq
\Box F_{ab} + 2 R_{a c b d} F^{cd} + \tensor{R}{_a ^c} F_{bc} - \tensor{R}{_b ^c} F_{a c} = 2 J_{[a ; b]} ,
\label{eq:waveF}
\eeq
In the absence of electromagnetic sources ($J_a = 0$), one may replace $F_{ab}$ with $\cF_{ab}$, if so desired. 
Alternatively, one may derive a wave equation for the vector potential,
\beq
\Box A^a - \tensor{R}{^a _b} A^b - \nabla^a \left( \tensor{A}{^b _{;b}} \right) = - J^a . \label{eq:waveA}
\eeq
The final term on the left-hand side is zero in Lorenz gauge, $\nabla_a A^a = 0$. 

The stress-energy due to the electromagnetic field, $T^{(F)}_{ab}$, is given by
\begin{eqnarray}
T^{(F)}_{ab} &\equiv& F_{ac} \tensor{F}{_b ^c} - \frac{1}{4} g_{ab} F_{cd} F^{cd}  \\
 &=& \frac{1}{2} \text{Re} \left( \tensor{\cF}{_a ^c} \cF^\ast_{bc} \right). \label{eq:stress-energy}
\end{eqnarray}
The stress-energy is traceless, $g^{ab} T^{(F)}_{ab} = \frac{1}{2} \cF_{ab} \cF^{ab\ast} = 0$, and it satisfies the conservation equation $\nabla_b T_{(F)}^{ab} = -\tensor{F}{^a _b} J^b$, which accounts for how energy is passed between the field and the charge distribution.

\subsubsection{Gravitational field\label{sec:GW}}
The spacetime metric $g_{ab}$ may be split into `background' $\hat{g}_{ab}$ and `perturbation' $h_{ab}$ parts, viz., $g_{ab} = \hat{g}_{ab} + \epsilon h_{ab}$, where $\epsilon$ is an order-counting parameter. In typical scenarios, one takes $\hat{g}_{ab}$ to describe a slowly-varying background geometry, with $h_{ab}$ a propagating part of small amplitude, such that terms at $O(\epsilon^2)$ in the field equations may be neglected. Under small changes in the coordinate system, $x^a \rightarrow x^{\prime a} = x^a + \epsilon \xi^a(x)$, where $\xi^a(x)$ is a vector field, the background $\hat{g}_{ab}$ remains unchanged and the metric perturbation undergoes a gauge transformation, $h_{ab} \rightarrow h^\prime_{ab} = h_{ab} - \mathcal{L}_\xi \hat{g}_{ab}$, where $\mathcal{L}_{\xi}$ is the Lie derivative along $\xi^a$. 

The Einstein field equation through linear order in $\epsilon$ yields the following,
\beq
\hat{\Box} \barh_{ab} + 2 \tensor{\hat{R}}{^c _a ^d _b} \barh_{cd} - 2 \left( \nabla_{(a} Z_{b)} - \frac{1}{2} g_{ab} \nabla_c Z^c \right) = -16 \pi G T^{(1)}_{ab} , \label{eq:boxh}
\eeq
where $\barh_{ab} = h_{ab} - \frac{1}{2}  g_{ab} h$ with $h\equiv \tensor{h}{^a _a}$, $Z_a \equiv \hat{\nabla}^b \barh_{ab}$, and $T_{ab}^{(1)}$ is the stress-energy at $O(\epsilon)$. Note that here covariant derivatives $\hat{\nabla}_a$ are defined with respect to $\hat{g}_{ab}$, and $\hat{g}^{ab}$ is used to raise indices. In some sense, the perturbation $h_{ab}$ propagates on a fixed background $\hat{g}_{ab}$. 

The term in parentheses in Eq.~(\ref{eq:boxh}) is zero in Lorenz gauge defined by $Z_a = \hat{\nabla}^b \barh_{ab} = 0$. We shall adopt this gauge in the next section.

The Riemann tensor can be uniquely decomposed into parts which are irreducible representations of the full Lorentz group \cite{Stephani:2003tm}: $R_{abcd} = C_{abcd} + E_{abcd} + G_{abcd}$, where $E_{abcd} = \frac{1}{2} \left(g_{ac}S_{bd} + g_{bd}S_{ac} - g_{ad}S_{bc} - g_{bc}S_{ad}\right)$, \, $G_{abcd} = \frac{1}{12}R\left(g_{ac}g_{bd}-g_{ad}g_{bc}\right)$ and $S_{ab} = R_{ab} - \frac{1}{4}Rg_{ab}$. 

The left and right duals are defined as follows,
\beq
{}^{\sim} C_{abcd} = \frac{1}{2} \varepsilon_{ab ef} \tensor{C}{^{ef} _{cd}} , \quad \quad
C^{\sim}_{abcd} = \frac{1}{2} \varepsilon_{cd ef} \tensor{C}{_{ab} ^{ef}} , 
\eeq
The Weyl tensor satisfies ${}^{\sim}C_{abcd} = C^{\sim}_{abcd}$. Thus we may introduce a complexified Weyl tensor
\beq
\mathcal{C}_{abcd} = C_{abcd} + i C^{\sim}_{abcd} ,
\eeq
which is self-dual with respect to both left and right duals. 

The Weyl tensor can be expressed in terms of derivatives of the Lanczos tensor $H_{abc}$, which in turn satisfies a wave equation $\Box H_{abc} = \mathcal{X}_{abc}$ \cite{Dolan:1994}, where the right-hand side depends on derivatives of the Ricci tensor and Ricci scalar; thus it is zero in a Ricci-flat spacetime. A confounding issue here is that the d'Alembertian operator $\Box = g^{ab} \nabla_a \nabla_b$ implicitly features the metric. An alternative approach is to take an additional derivative of the Bianchi identity, to obtain a wave equation for the Riemann tensor. In a Ricci-flat spacetime, 
\beq
\Box R_{abcd} + 2 R_{aecf} \tensor{R}{_b ^e _d ^f} - 2 R_{aedf} \tensor{R}{_b ^e _c ^f} + R_{abef} \tensor{R}{_{cd} ^{ef}}  = 0 .
\eeq

\section{Geometrical optics\label{sec:go}}

Let us suppose now that the wavelength is short in comparison to all other relevant length scales; and that the inverse frequency is short in comparison to other relevant timescales.

\subsection{Scalar field\label{sec:go-scalar}}
We begin with a \emph{geometrical-optics ansatz},
\beq
\Phi(x) = \cA(x) \exp\left( i \omega \Psi(x) \right) . \label{eq:ansatz}
\eeq
Here $\omega$ serves as an order-counting parameter, and the real fields $\Psi(x)$ and $\cA(x)$ are the phase and amplitude, respectively. Loosely, we may call $\omega$ the `frequency', but with the note of caution that an observer with unit tangent vector $u^a$ would actually measure a wave frequency of $-\omega u^a \nabla_a \Psi$.

Substituting (\ref{eq:ansatz}) into the wave equation (\ref{eq:kg}) leads to 
\beq
\left( - \omega^2  k^a k_a \cA + i \omega \left( k^a \nabla_a \cA + \nabla_a k^a \cA \right) + \Box \cA \right) e^{i \omega \Psi} = 0 ,
\eeq
where $k_a \equiv \nabla_a \Psi$. 
The standard geometrical-optics approach is to examine this equation order-by-order in $\omega$. At $O(\omega^2)$, $k^a k_a = 0$, thus the tangent vector is null. As $k_a$ is a gradient, it follows as a consequence that it satisfies the geodesic equation,
\beq
\frac{D k_a}{d \nu} \equiv k^b \nabla_b k_{a} = 0,
\eeq
where $\nu$ is an affine parameter. The integral curves of $k^a$, that is, the spacetime paths $x^a(\nu)$ satisfying $\frac{dx^a}{d\nu} = k^a$, are null geodesics which lie in the hypersurface of constant phase ($\Psi(x) = \text{constant}$). These are known as the null generators, and they may be found from the constrained Hamiltonian
$\mathcal{H}[x^a, k_a] = \frac{1}{2} g^{a b}(x) k_{a} k_{b}$, 
where $\mathcal{H} = 0$ and $k_{a} \equiv g_{a b} \frac{dx^b}{d\nu}$.

At $O(\omega^{1})$, one obtains a transport equation for the amplitude, 
\beq
k^a \nabla_a \cA = -\frac{1}{2} \vartheta \cA, \label{eq:amplitude}
\eeq
where $\vartheta \equiv \nabla_a k^a$ is the expansion scalar. The transport equation for the amplitude $\cA$ ensures the conservation of flux, $\nabla_a \left( \cA^2 k^a \right) = 0$.

Typically, the last equation $\Box \cA = 0$, at $O(\omega^0)$, is discarded, meaning that we have only an approximate solution. (If one could find a solution for $\cA$ that was consistent with the equations above, then one would have an exact solution; however $\Box \cA = 0$ appears no easier to solve than the original equation $\Box \Phi = 0$, it would seem). 

Inserting Eq.~(\ref{eq:ansatz}) into Eq.~(\ref{eq:scalar-se}) gives the stress-energy,
\beq
T^{(\Phi)}_{ab} = \omega^2 \cA^2 k_a k_b  + \omega^0 \left(\nabla_a \cA \nabla_b \cA - \frac{1}{2} g_{ab} \nabla_c \cA \nabla^c \cA \right) .
\eeq
At order $\omega^2$ the stress-energy has the form of a null fluid. There is no piece at $O(\omega^1)$. 

It is straightforward to show that $\nabla^b T^{(\Phi)}_{ab} = \omega^0 \Box \cA = 0$ by using flux conservation ($\nabla_a \left( \cA^2 k^a \right) = 0$).

\subsection{Electromagnetic field\label{sec:go-em}}

A standard approach is to introduce a geometrical-optics ansatz for the vector potential $A^a$ into the wave equation (\ref{eq:waveA}) and to adopt Lorenz gauge ($\nabla_a A^a = 0$); see for example Box 5.6 in Ref.~\refcite{Poisson:Will:2014}. 

Another approach \cite{Kristian:1965sz, Ehlers:1967, Anile:1976gq}, which we follow here, is to introduce an ansatz for the Faraday tensor $F_{ab}$ itself. This helps to expedite the stress-energy tensor calculation, and removes any possible doubt about the gauge invariance of the results obtained. 
We take the \emph{ansatz},
\beq
\cF_{ab} = \cA(x) \ff_{ab}(x) \exp\left( i \omega \Psi(x) \right) .   \label{eq:Fansatz}
\eeq 
Here $\ff_{ab}$ is the polarization bivector, a self-dual bivector field ($\ff_{ab} = -\ff_{ba}$, $\widetilde{\ff}_{ab} = - i \ff_{ab}$). 

In a charge-free region ($J^a = 0$), inserting (\ref{eq:Fansatz}) into the wave equation (\ref{eq:waveF}) yields
\beq
-\omega^2 k^c k_c \cA \ff_{ab}+  i \omega \left[\left(2 k^c \nabla_c \cA + \left(\nabla_c k^c\right) \cA \right) \ff_{ab} +  \cA k^c \nabla_c \ff_{ab} \right] + O(\omega^0) = 0,
\eeq
Once again, the gradient of the phase is null, $k_a k^a = 0$ and thus tangent to a null generator. 
We shall choose to split the amplitude $\cA$ and the polarization $\ff_{ab}$ such that, once again, $k^a \nabla_a \cA = -\frac{1}{2} \vartheta \cA$, and now $k^c \nabla_c \ff_{ab} = 0$. Thus, the flux is once again conserved, $\nabla_a \left( \cA^2 k^a \right) = 0$, and the polarization bivector is parallel-propagated along the null generator: $k^c \nabla_c \ff_{ab} = 0$. This is basis for the Skrotskii/Rytov effect, or gravitational Faraday rotation.  

At leading order the polarization bivector is transverse, $\ff_{ab} k^b = 0$. This follows from $\nabla_b \tensor{\cF}{^{ab}} = 0$ at $O(\omega^1)$.

\subsubsection{Circular and linear polarizations}
A circularly-polarized wave satisfying $\ff_{ab} k^b = 0$ and $k^c \nabla_c \ff_{ab}$ is constructed by choosing
\beq
f_{ab} = 2 k_{[a} m_{b]}, \label{eq:f0}
\eeq
where $m^a$ is any complex null vector satisfying $m_a m^a = m_a k^a = 0$ and $m_a \mbar^a = 1$ (where $\mbar^a$ is the complex conjugate of $m^a$), that satisfies
\beq
k^b \nabla_b m^a = \alpha(\nu) k^a,
\eeq
where $\alpha(\nu)$ is any scalar function. With the choice $\alpha = 0$, the vector $m^a$ is parallel-propagated along the null generator, $k^b \nabla_b m^a = 0$. 

Typically, $m^a$ is constructed from a pair of legs from an orthonormal triad, e.g.~$m^a = \frac{1}{\sqrt{2}} \left( e_1^a + i e_2^a \right)$, and conversely, $e_1^a = \frac{1}{\sqrt{2}} \left( \mbar^a + m^a \right)$ and $e_2^a = \frac{1}{\sqrt{2}}  \left( \mbar^a - m^a \right)$. The handedness depends on the sign of $\omega$ and the handedness of $m_a$. Henceforth, we shall assume that $m_a$ is constructed such that $i \varepsilon_{a b c d} u^a k^b m^c \mbar^d$ is positive for any future-pointing timelike vector $u^a$. The wave is right-hand polarized (left-hand polarized) if the frequency $\omega$ is positive (negative).

A linear polarization may be constructed by adding together left- and right-handed polarizations, e.g.,
$$
\cF_{ab} = 2 \cA k_{[a} m_{b]} e^{i \kappa_1} \cos\left( \omega \Psi + \kappa_2 \right),
$$
where $\kappa_1$ and $\kappa_2$ are constants. 

\subsubsection{Stress-energy}
The circular and linearly-polarized fields are null at leading order in $\omega$. This can be seen by 
inserting Eq.~(\ref{eq:ansatz}) into Eq.~(\ref{eq:frame-inv}) to obtain $\Upsilon = 0$, after noting that $\ff_{ab} \ff^{ab} = 0$ for circularly-polarized waves. 

Inserting Eq.~(\ref{eq:Fansatz}) into Eq.~(\ref{eq:stress-energy}) gives a leading-order (in $\omega$) result for the stress-energy,
\beq
T_{ab} = \frac{1}{2} \cA^2 k_a k_b + O(\omega^{-1}) .
\eeq
Again, the stress-energy has the form of a null fluid at leading order. It is straightforward to show that $\nabla^b T_{ab} = 0$ at this order by using flux conservation ($\nabla_a \left( \cA^2 k^a \right) = 0$). The sub-dominant corrections to the stress-energy at $O(\omega^{-1})$ and beyond are examined in Ref.~\refcite{Dolan:2018ydp}.
 
\subsection{Gravitational field\label{sec:go-gw}}
Following the pattern of the previous sections, let
\beq
\barh_{ab} = \text{Re} \left( \cA \hh_{ab} e^{i \omega \Psi} \right) ,
\eeq
where $\hh_{ab}$ is the polarization. From the Lorenz-gauge condition at leading order, one obtains the transversality condition $\hh_{ab} k^b = 0$. From the wave equation (\ref{eq:boxh}), one obtains
\beq
-\omega^2 k^c k_c \hh_{ab}+  i \omega \left[\left(2 k^c \nabla_c \cA + \left(\nabla_c k^c\right) \cA \right) \hh_{ab} +  \cA k^c \nabla_c \hh_{ab} \right] + O(\omega^0) = 0,
\eeq
Following the argument of the previous section, we have once again $k^a \nabla_a \cA = -\frac{1}{2} \vartheta \cA$, and the parallel transport of the polarization, viz., $k^c \nabla_c \hh_{ab} = 0$.

The Lorenz gauge leaves a residual freedom; one may add any gauge perturbation with a gauge vector $\xi_a$ satisfying $\Box \xi_a = 0$. A circularly-polarized wave may be constructed by choosing
$
\hh_{ab} = m_{(a} m_{b)}
$, where $m_a$ is any complex null vector satisfying $m_a k^a = m_a m^a = 0$, $m_a \mbar^a = 1$ and the parallel-transport equation $k^b \nabla_b m_a = 0$.

\section{Gravitational lensing\label{sec:lensing}}

In this section we consider at how the geometrical-optics approximation informs the theory of gravitational lensing \cite{Perlick:2004tq, Seitz:1994xf}. The intuitive notion of a beam of neighbouring rays is described by a null congruence. We consider below a parallel-propagated null tetrad defined on a congruence.

\subsection{Geodesic deviation\label{sec:geodesic-deviation}}
 
 \subsubsection{A null tetrad}
In Secs.~\ref{sec:go-scalar} and \ref{sec:go-em} we introduced the (future-pointing) null tangent vector $k^a$ and a complex null vector $m^a = \frac{1}{\sqrt{2}} \left( e_1^a + i e_2^a \right)$. We may complete the null tetrad by introducing an auxiliary null vector $n^a$ \cite{Poisson:2004}: a future-pointing null vector field satisfying $k_a n^a = -1$ and $m_a n^a = 0$, such that
\beq
\varepsilon^{abcd} = i 4! k^{[a} n^{b} m^{c} \mbar^{d]} .
\eeq
The metric is $g^{ab} = -2k^{(a} n^{b)} + 2 m^{(a} \mbar^{b)}$. 


%

 \subsubsection{Geodesic deviation\label{subsec:geodesic-deviation}}
Consider two neighbouring geodesics (null, spacelike or timelike), $\gamma_0$ and $\gamma_1$, with spacetime paths $x_0^a(\nu)$ and $x_1^a(\nu)$ \cite{Poisson:2004} with $\nu$ an affine parameter. Between $\gamma_0$ and $\gamma_1$, introduce a one-parameter family of null geodesics $x^a(\nu,s)$, such that $x_0^a(\nu) = x^a(\nu,0)$ and $x_1^a(v)=x^a(\nu,1)$. The vector field $u^a \equiv \partial x^a / \partial \nu$ is tangent to the geodesics, and thus satisfies $u^b \nabla_b u^a = 0$. The vector field $\xi^a \equiv \partial x^a / \partial s$ spans the family, though it is not tangent to a geodesic, in general. The identity $\partial \xi^a / \partial \nu - \partial u^a / \partial s = 0$ (partial derivatives commute) implies that $\xi^a$ is Lie-transported along each geodesic, $\mathcal{L}_u \xi^a \equiv u^b \nabla_b \xi^a - \xi^b \nabla_b u^a = 0$. An elementary consequence is that $\frac{d}{d\nu} \left( \xi^a u_a \right) = 0$, and so $\xi^a u_a$ is constant along each geodesic.  A standard calculation \cite{Poisson:2004} shows that the acceleration of the deviation vector $\xi^a$ is given by
\begin{eqnarray}
\frac{D^2 \xi^a}{d\nu^2} &\equiv& u^c \nabla_c \left( u^b \nabla_b \xi^a \right)  \nonumber \\
 &=& -\tensor{R}{^a _{bcd}} u^b \xi^c u^d .  \label{eq:geodesic-deviation}
\end{eqnarray}
This is the geodesic deviation equation, which describes how spacetime curvature leads to a relative acceleration between neighbouring geodesics, even if they start out parallel \cite{Poisson:2004}. 

Henceforth we examine the null case with $u^a = k^a$. We may express the deviation vector, restricted to a central null geodesic $\xi^a(\nu) = \partial x / \partial s |_{s=0}$, in terms of the null basis on that geodesic. Let 
\beq \xi^a = a(\nu) k^a + b(\nu) n^a + \overline{z}(\nu) m^a + z(\nu) \bar{m}^a ,
\eeq 
where $a$ and $b$ are real and $z$ is complex. After inserting into Eq.~(\ref{eq:geodesic-deviation}) and projecting onto the tetrad, one obtains a hierarchical system of equations:
\begin{subequations}
\begin{eqnarray}
\ddot{b} &=& 0 , \label{eq:bddot} \\
\ddot{a} &=& b R_{nknk} + \overline{z} R_{knkm} + z R_{knk\mbar} , \label{eq:addot} \\
\ddot{z} &=& -b R_{kmkn} - \overline{z} R_{kmkm} - z R_{kmk\mbar} ,  \label{eq:zddot}
\end{eqnarray}
\end{subequations}
where $\ddot{a} \equiv d^2 a / d\nu^2$, etc., and $R_{k\mbar km} \equiv R_{abcd} k^a \mbar^b k^c m^d$, etc. Note that Eq.~(\ref{eq:bddot}) is consistent with $b = -\xi^a k_a = \text{const.}$, as established above. If one sets $b=0$  then 
\begin{subequations}
\label{eq:azdot}
\begin{eqnarray}
\ddot{a} &=& \left(\Phi_{00} + \Psi_1 \right) \overline{z} + \left(\overline{\Phi}_{00} + \overline{\Psi}_1 \right) z, \\
\ddot{z} &=& - \Phi_{00} z -  \Psi_0 \overline{z} , 
\end{eqnarray}
\label{eq:zdeviation}
\end{subequations}
where the Ricci and Weyl scalars are given by $\Phi_{00} = \frac{1}{2} R_{kk} = R_{kmk\mbar}$, 
$\Psi_0 = C_{kmkm} = R_{kmkm}$ and $\Psi_1 = C_{knkm}$ (here $C_{kmkm} = C_{abcd} k^a m^b k^c m^d$, etc.). 

\subsubsection{Beam cross section}
The complex value $z = \frac{1}{\sqrt{2}} \left(x + iy\right)$ corresponds to a point $(x,y)$ on the wavefront with position vector $\hat{\xi}^a = \overline{z} m^a + z \mbar^a$, with $m^a = \frac{1}{\sqrt{2}} \left( e_1^a + e_2^a \right)$, where $e_i^a$ are orthogonal unit vectors. If $z_1$ and $z_2$ are any pair of linearly-independent solutions of Eq.~(\ref{eq:zdeviation}) then $z(\phi) = \cos \phi \, z_1 + \sin \phi \, z_2$, where $\phi \in [-\pi, \pi)$, corresponds to an ellipse in the wavefront. One may show that the principle axes are given by $z_+ = \cos \phi_0 \, z_1 + \sin \phi_0 \, z_2$ and $z_- = -\sin \phi_0 \, z_1 + \cos \phi_0 \, z_2$, where $\tan(2 \phi_0) = 2 \text{Re}(z_1 \overline{z}_2) / \left( |z_1|^2 - |z_2|^2 \right)$. The semi-major/semi-minor axes $d_\pm = \sqrt{2} |z_\pm|$ are given by $d_+ d_- = 2 \left| \text{Im} ( z_1 \overline{z}_2 ) \right|$ and $d_+^2 + d_-^2 = 2 \left( |z_1|^2 + |z_2|^2 \right)$. It follows that the cross sectional area $A = \pi d_+ d_-$ satisfies the transport equation $\dot{A} = \left(\varrho + \overline{\varrho} \right) A$. Comparing this with Eq.~(\ref{eq:amplitude}) shows that the square of the wave amplitude, $\mathcal{A}^2$, scales in proportion to the inverse of the cross-sectional area of the beam, $A^{-1}$. 

\subsection{Transport equations for optical \& Newman-Penrose scalars\label{subsec:optical-scalars}}
 \subsubsection{Sachs' equations}
Now let $\dot{z} = -\rho(\nu) z - \sigma(\nu) \overline{z}$, where $\rho(\nu)$ and $\sigma(\nu)$ are complex functions. From $\mathcal{L}_k \xi^a = 0$ and $b=0$, it follows that $\rho = -m^a k_{a;b} \mbar^b$ and $\sigma = -m^a k_{a;b} m^b$. (The signs have been chosen here so that $\rho$ and $\sigma$ are consistent with the usual Newman-Penrose definitions \cite{Newman:1961qr}). Inserting into Eq.~(\ref{eq:zdeviation}) and equating the coefficients of $z$ and $\overline{z}$ leads to a pair of first-order transport equations,
\begin{eqnarray}
\dot{\rho} &=& \rho^2 + \sigma \overline{\sigma} + \Phi_{00} , \\
\dot{\sigma} &=& \sigma \left( \rho + \overline{\rho} \right) + \Psi_0 .
\end{eqnarray}
These are known as the Sachs equations \cite{Perlick:2004tq}. The real and imaginary parts of $\rho$ and $\sigma$ yield the \emph{optical scalars} \cite{Jordan:2013, Kantowski:1968, Frolov:1998wf}: $-\rho = \theta + i \varpi$, $-\sigma = \varsigma_1 + i \varsigma_2$, where $\theta = \frac{1}{2} \tensor{k}{^a _{;a}}$, $\varpi$ and $(\varsigma_1, \varsigma_2)$ are known as the expansion, twist and shear, respectively. The twist $\varpi$ is zero for a hypersurface-orthogonal congruence; therefore it is zero for a congruence lying in a constant-phase hypersurface, such as those appearing in geometrical-optics, because $\nabla_{[a} k_{b]} = \nabla_{[a} \nabla_{b]} \Psi = 0$. Kantowski \cite{Kantowski:1968} proved that a (2D) wavefront seen by an observer with tangent vector $u^a$ has principal curvatures $\kappa_{\pm}$ given by $\kappa_{\pm} = (-u^a k_a)^{-1} \left( \theta \pm |\sigma| \right)$.

A shortcoming of the Sachs equations is that the quantities $\rho$ and $\sigma$ necessarily diverge at a conjugate point, where neighbouring rays cross. By contrast, the second-order equation (\ref{eq:zdeviation}) does not suffer from divergences. The quantities $\rho$ and $\sigma$ can be found from any linearly-independent pair of solutions of Eq.~(\ref{eq:zdeviation}), $z_1$ and $z_2$, by solving
\beq
\begin{pmatrix} \dot{z}_1 \\ \dot{z}_2 \end{pmatrix} = -
\begin{pmatrix} z_1 & \overline{z}_1 \\ z_2 & \overline{z}_2 \end{pmatrix} 
\begin{pmatrix} \rho \\ \sigma \end{pmatrix} . \label{eq:optical-z1z2}
\eeq
The inversion breaks down wherever $\text{Im} \left( z_1 \overline{z}_2 \right) = 0$, i.e., at conjugate points. However, note that $z_1$ and $z_2$ themselves are regular at conjugate points. Thus, by solving second-order equations for $z$, rather than the first-order Sachs' equations, we can track $\rho$ and $\sigma$ through conjugate points.

\subsubsection{Newman-Penrose scalars}
There are further scalar quantities associated with a null congruence. Newman and Penrose \cite{Newman:1961qr} introduced a set of scalars, defined in terms of projections of first derivatives of the null tetrad legs. One is trivially zero for a geodesic congruence: $\kappa \equiv m^a k_{a;b} k^b = 0$. Two further scalars are zero for a parallel-propagated tetrad: $\pi \equiv \overline{m}^a k^b \nabla_b n_a = 0$ and $\epsilon \equiv -\frac{1}{2}\left(n^a k^b \nabla_b k_a - \overline{m}^a k^b \nabla_b m_a \right)$. Six further scalars, of relevance to a twist-free geodesic congruence, are defined below:
\begin{subequations}
\label{eq:NP}
\begin{align}
\sigma &= -m^a k_{a;b} m^b  ,&
\tau &= - m^a k_{a ; b} n^b , \\
\rho &= -m^a k_{a ; b} \mbar^b , & 
\chi &=  \mbar^a m_{a ; b} m^b , \\
\mu &= \mbar^a n_{a;b} m^b, &
\lambda &= \mbar^a n_{a;b} \mbar^b ,
\end{align}
\end{subequations}
where the semi-colon denotes the covariant derivative ($k_{a;b} \equiv \nabla_b k_a$, etc.).

Some identities follow from applying $g^{ab} = -k^a n^b - n^a k^b + m^a \mbar^b + \mbar^a m^b$ together with the fact that $k_a$ is a gradient, $k_{[a;b]} = 0$. For example, $\rho$ is purely real due the twist-free (gradient) property of the null tetrad, and $\rho = -\frac{1}{2} \vartheta$ where $\vartheta = \tensor{k}{^a _{;a}}$ is the expansion scalar \cite{Poisson:2004}. Furthermore, $\tau = \beta + \bar{\alpha}$ in the twist-free case, where $\alpha = \frac{1}{2}\left(k^a n_{a;b} \mbar^b - m^a \mbar_{a;b} \mbar^b \right)$ and $\beta = \frac{1}{2} \left(\mbar^a m_{a;b} m^b - n^a k_{a;b} m^b \right)$ are further Newman-Penrose scalars. Instead of using $\alpha$ and $\beta$, I have introduced a new symbol, $\chi \equiv \beta - \overline{\alpha}$. 

By noting that $\dot{a} = -n^c k^b \nabla_b \xi^c = \overline{z} \tau + z \overline{\tau}$, one can find the Newman-Penrose quantity $\tau$ from a pair of solutions of Eq.~(\ref{eq:zdeviation}) by solving
\beq
\begin{pmatrix} \dot{a}_1 \\ \dot{a}_2 \end{pmatrix} = 
\begin{pmatrix} z_1 & \overline{z}_1 \\ z_2 & \overline{z}_2 \end{pmatrix} 
\begin{pmatrix} \overline{\tau} \\ \tau \end{pmatrix} . \label{eq:tau}
\eeq
The quantities $\chi$, $\mu$ and $\lambda$ can be found by considering differential precession, as we show below.

\subsubsection{Differential precession}
Consider a congruence of null geodesics (see Sec.~\ref{subsec:geodesic-deviation}) with a 2D cross section seen by an observer with tangent vector $u^a$ and worldline $\gamma$. The cross section (i.e.~the 2D instantaneous wavefront) is spanned by a basis $m^a = \frac{1}{\sqrt{2}} \left( \hat{e}_1^a + i \hat{e}_2^a \right)$ and $\mbar^a$, such that $k^a m_a = u^a m_a = 0$ and $m^a \mbar_a = 1$. It is natural for an observer to choose a basis that is `straight' in their vicinity, in the sense that $\left. \hat{\xi}^b \nabla_b m^a \right|_\gamma = 0$ for any $\hat{\xi}^a \equiv \overline{z} m^a + z \overline{m}^a$. 
However, a basis that starts out straight does not remain straight, in general, once it is parallel-propagated along the rays in a geodesic null congruence in the presence of a gravitational field. (See e.g.~Ref.~\refcite{Nichols:2011pu} for a discussion of differential precession along \emph{timelike} geodesics.)

Let $\zeta^a \equiv \left. \xi^b \nabla_b m^a \right|_\gamma$, where $\mathcal{L}_k \xi^a = 0$ and $k^b \nabla_b k_{a} = 0$. One may follow steps analogous to those in the derivation of the geodesic deviation equation, Eq.~(\ref{eq:geodesic-deviation}), to derive the \emph{differential precession equation},
\beq
\frac{D \zeta^a}{d v} = -\tensor{R}{^a _{bcd}} m^b \xi^c k^d . 
\eeq
Decomposing in the null tetrad, $\zeta^a = \alpha k^a + \dot{z} n^a + \eta m^a$, where $\alpha$ and $\eta$ are  functions of the affine parameter, leads to
\begin{subequations}
\label{eq:alphaeta}
\begin{eqnarray}
\dot{\alpha} &=& z \overline{\Psi}_2 , \\
\dot{\eta} &=&  \overline{z} \Psi_1 - z \overline{\Psi}_1.  \label{eq:eta-transport}
\end{eqnarray}
\end{subequations}
in a Ricci-flat spacetime, where $\alpha = \overline{\mu} z + \overline{\lambda} \overline{z}$, $\eta = \overline{z} \chi - z \overline{\chi}$, and $\mu$, $\lambda$ and $\chi$ are Newman-Penrose scalars, and $\Psi_2 = C_{abcd} k^a m^b \mbar^c n^d$. These scalars can be found from any pair of linearly-independent solutions $(c_1, \alpha_1, \eta_1)$ and $(c_2, \alpha_2, \eta_2)$ satisfying Eqs.~(\ref{eq:zdeviation}) and (\ref{eq:eta-transport}), by inverting 
\beq
\begin{pmatrix} \eta_1 \\ \eta_2 \end{pmatrix} = \begin{pmatrix} \overline{z}_1 & z_1 \\ \overline{z}_2 & z_2 \end{pmatrix} \begin{pmatrix} \chi \\ -\overline{\chi} \end{pmatrix} \quad \text{and} \quad
\begin{pmatrix} \alpha_1 \\ \alpha_2 \end{pmatrix} = \begin{pmatrix} \overline{z}_1 & z_1 \\ \overline{z}_2 & z_2 \end{pmatrix} \begin{pmatrix} \overline{\mu} \\ \overline{\lambda} \end{pmatrix}.  \label{eq:chimulambda}
\eeq
As for Eq.~(\ref{eq:optical-z1z2}), this procedure fails at a conjugate point. On the other hand, the functions $\alpha$ and $\eta$ are regular there, and so it is straightforward to move beyond the conjugate point.

Taken together, Eqs.~(\ref{eq:azdot}), (\ref{eq:optical-z1z2}), (\ref{eq:tau}), (\ref{eq:alphaeta}) and (\ref{eq:chimulambda}) comprise a practical method, based around transport equations, for calculating the Newman-Penrose scalars $\rho$, $\sigma$, $\tau$, $\chi$, $\mu$ and $\lambda$ for a null tetrad that is parallel-propagated along a null geodesic. By contrast, the traditional transport equations of Newman-Penrose, given below, run into difficulties at the first conjugate point, where the Newman-Penrose scalars diverge:
\begin{subequations} \label{eqs:NPtransport}
\begin{eqnarray}
\dot{\rho} &=& \rho^2 + \sigma \overline{\sigma} , \\
\dot{\sigma} &=& 2 \rho \sigma + \Psi_0 , \\
\dot{\chi} &=& \rho \chi -\sigma \overline{\chi} + \Psi_1 , \\
\dot{\tau} &=& \rho \tau + \sigma \overline{\tau} + \Psi_1 , \\
\dot{\lambda} &=& \rho \lambda + \overline{\sigma} \mu , \\
\dot{\mu} &=& \rho \mu + \sigma \lambda + \Psi_2 .
\end{eqnarray}
\end{subequations}

\subsubsection{Asymptotics}
In a flat (Minkowki) region of spacetime, the transport equations have exact solutions. 
A general pair of solutions to $\ddot{z} = 0$ such that $\rho$ is real are $z_1 = C_1 e^{i\phi_1} (t + \alpha)$ and $z_2 = C_2 e^{i \phi_2} (t + \overline{\alpha})$, where $C_i$, $\phi$ are real constants and $\alpha$ is a complex constant. Without loss of generality for describing the congruence, we choose $C_1 = C_2$ and $e^{i \phi_2} = i e^{i \phi_1} = e^{i (\phi + \pi /4)}$. Solving (\ref{eq:optical-z1z2}) gives
\begin{align}
\rho &= -\frac{1}{2} \left[ (\nu + a + b)^{-1} + (\nu + a - b)^{-1} \right] &&= - \nu^{-1} + a \nu^{-2} - (a^2+b^2) \nu^{-3} + \ldots  \\
\sigma &= \frac{e^{2i\phi}}{2} \left[ (\nu + a + b)^{-1} - (\nu + a - b)^{-1} \right] &&= - e^{2 i \phi} \left(b \nu^{-2} - 2ab \nu^{-3} + \ldots \right) ,
\end{align}
where $a \equiv \text{Re}(\alpha)$ and $b = \text{Im}(\alpha)$. 

By inspection of Eq.~(\ref{eqs:NPtransport}), we can deduce that, in the limit $\nu \rightarrow \infty$, the Newman-Penrose coefficients $\rho$, $\chi$, $\tau$, $\lambda$ and $\mu$ decay as $O(\nu^{-1})$; and $\sigma$ decays as $O(\nu^{-2})$.

\subsection{Modified phase\label{subsec:diffprecession}}

In this section we argue that differential precession of the null tetrad across a beam can be interpreted as an additional phase term in the leading-order geometrical-optics expansion. The gradient of that phase can be tentatively interpreted as a spin-deviation contribution to the tangent vector $k^a$ at order $\omega^{-1}$, whose sign depends on the handedness of the polarization.  

\begin{figure}
\begin{minipage}[c]{13cm}
 \begin{center}
  \includegraphics[align=c,width=5cm]{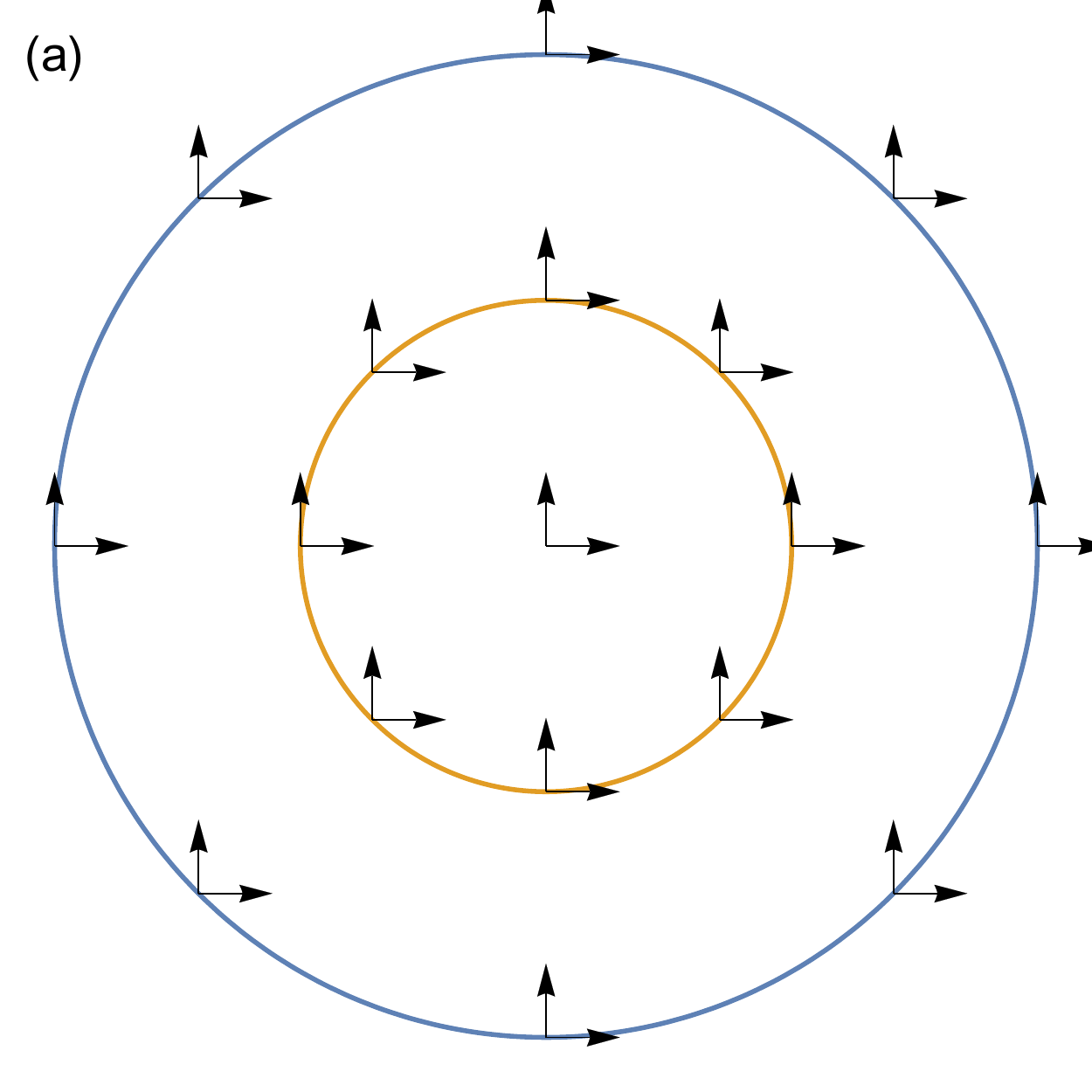}  \; \quad \;
  \includegraphics[align=c,width=4cm]{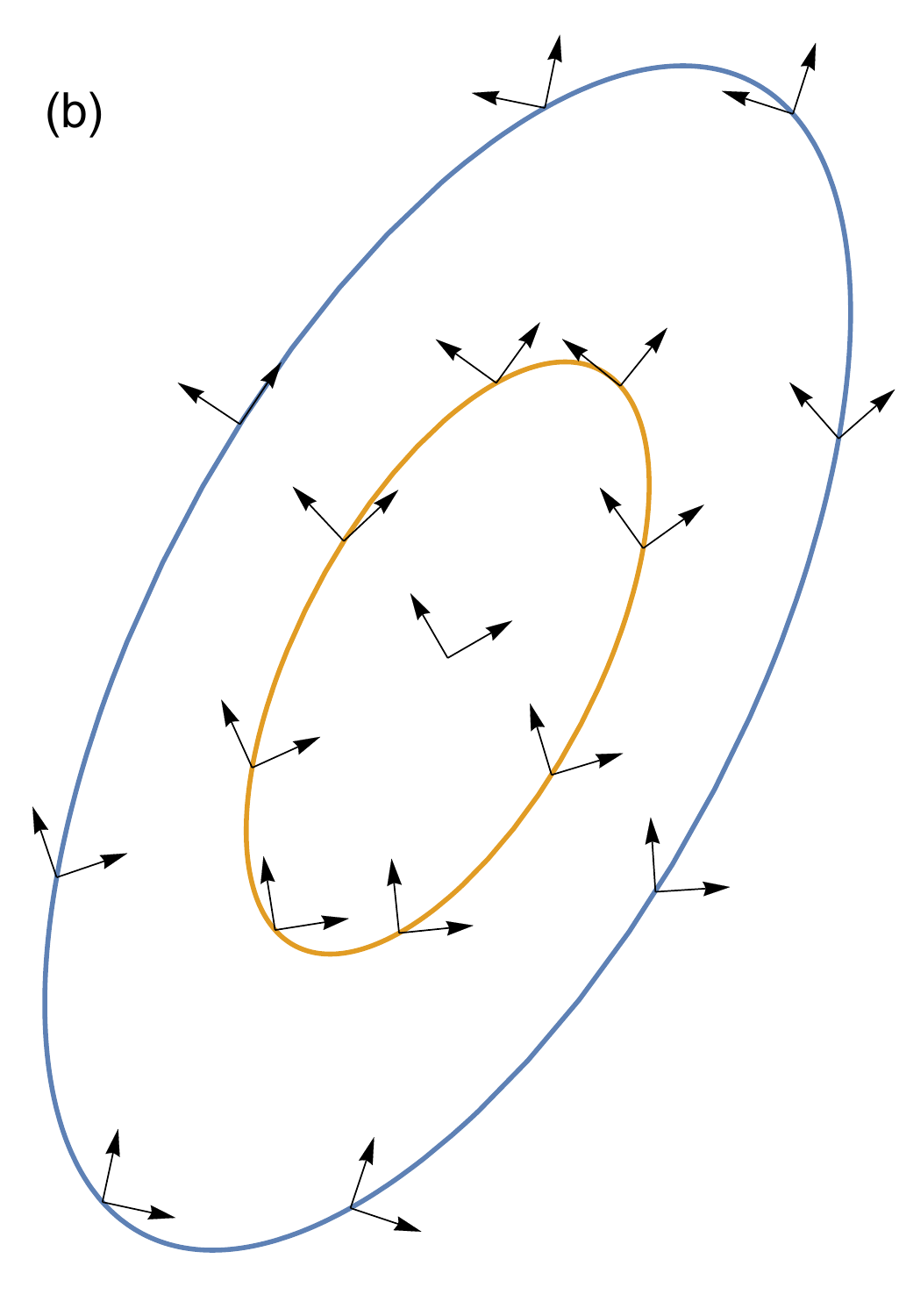}
 \end{center}
\end{minipage}
 \caption{(a) Circular cross section of a beam of rays with a basis $m^a = \frac{1}{\sqrt{2}} \left(e_1^a + i e_2^b\right)$ that is `straight': $\mbar^b \nabla_b m^a = 0$.  (b) Elliptical cross section of the same beam after propagating through a gravitational field.  The parallel-propagated basis $m^a$ has undergone differential precession such that $\mbar^b \nabla_b m^a \neq 0$.}
 \label{fig:differential-precession}
\end{figure}

Suppose that the cross section of the congruence is initially circular and the frame $m^a$ is initially `straight', as shown in Fig.~\ref{fig:differential-precession}(a). After the congruence has passed through a gravitational field, the cross section will be elliptical, in general; furthermore, the basis will \emph{not} be straight, as shown in Fig.~\ref{fig:differential-precession}(b), due to differential precession. An observer with tangent vector $U^a = \frac{1}{2\beta} k^a + \beta n^a$ (where $\beta > 0$ is a free parameter) will see a wavefront spanned by $m^a$ and $\overline{m}^a$. However, that observer would naturally prefer a `straight' basis $\hat{m}^a = e^{-i \varphi} m^a$, such that $\hat{\xi}^b \nabla_b \hat{m}^a = 0$, where the gradient of the phase is
\beq
m^a \nabla_a \varphi = - i \chi ,
\eeq
with $\chi$ defined in Eq.~(\ref{eq:NP}).

Now consider the geometric optics solution for the Faraday tensor, Eqs.~(\ref{eq:Fansatz}) \& (\ref{eq:f0}), from the perspective of this observer. With the observer's preference for a locally-straight basis $\hat{m}^a = e^{-i\varphi} m^a$, one could write
\beq
\cF_{ab} = 2 k_{[a} \hat{m}_{b]} \mathcal{A} \exp\left(i \omega \Psi^\prime \right), \quad \quad \Psi^\prime \equiv \Psi + \omega^{-1} \varphi .
\eeq 
The gradient of the \emph{modified phase} $\Psi^\prime$ is
\beq
K_a \equiv \nabla_a \Psi^\prime = k_a + \omega^{-1} \left( i \overline{\chi} m_a - i \chi \mbar_a  + v_a \right) ,
\label{eq:Kfirst}
\eeq
where $v_a m^a = v_a \overline{m}^a = 0$. It is tempting to interpret $K_a$ as an `effective' tangent vector which accounts for the effect of differential precession. Going one step further, we note that one could introduce $M_a \equiv e^{i \varphi} \left(m_a - \omega^{-1} i \chi n_a \right)$ such that  $K^a M_a = 0$, leading to $2 K_{[a} M_{b]} = U_{ab} + i \chi \omega^{-1} W_{ab}$. 

It is important to note that the vectors $K_a$ and $M_a$ depend on the choice of null basis, and they are \emph{not} invariant under the parabolic Lorentz transformation $k^{\prime a} = k^a$, $m^{\prime a} = m^a - B k^a$, $n^{\prime a} = n^a - B m^a - \overline{B} \mbar^a + B \overline{B} k^a$, where $B$ is a complex scalar field.

\section{Discussion and conclusions\label{sec:discussion}}

Here, we have reviewed the geometrical-optics method and its application to wave equations describing the propagation of scalar, electromagnetic and gravitational waves in a curved spacetime. The method reduces the problem of solving hyperbolic PDEs to that of solving transport equations (ODEs) along rays, which are the null generators of hypersurfaces of constant phase. In the electromagnetic and gravitational cases, the polarization is parallel-propagated along these rays. This led us on to consider a null tetrad that is parallel-propagated along a ray (i.e.~a Sachs basis), and the associated optical scalars and a subset of Newman-Penrose scalars.

The cross section of a beam of rays passing through a gravitational field is distorted from a circle to an ellipse, in general (see Fig.~\ref{fig:differential-precession}). The square of the wave amplitude varies in inverse proportional to the cross-sectional area of the beam. At conjugate points, where neighbouring rays cross, the ellipse has zero area -- and thus the geometrical-optics amplitude diverges. The standard transport equations in first-order form -- i.e.~Sachs equations and Newman-Penrose transport equations -- break down at the first conjugate point, where the optical scalars diverge.

Here we have presented an alternative method, starting from the geodesic deviation equation, which mixes second order ODEs and first-order ODEs. The optical scalars ($\rho, \sigma$) can be obtained from a pair of complex scalars $z_1$ and $z_2$ which represent deviation vectors in the beam. Note that $z_1$ and $z_2$ are regular at a conjugate point; it is the loss of linear independence in $z_1$ and $z_2$ that causes the divergence of the optical scalars. 

Taken together, Eqs.~(\ref{eq:azdot}), (\ref{eq:optical-z1z2}), (\ref{eq:tau}), (\ref{eq:alphaeta}) and (\ref{eq:chimulambda}) comprise a practical method for calculating the NP scalars $\rho$, $\sigma$, $\tau$, $\chi$, $\mu$ and $\lambda$ along an entire ray. The method does not break down at the first conjugate point encountered; it is merely the case the quantities are ill-defined there. 

The apparent `twisting' in the polarization of an electromagnetic or gravitational wave due to parallel-propagation in a curved spacetime goes by the name of gravitational Faraday rotation, or the Skrotskii/Rytov effect. Here we have also considered \emph{differential precession}, i.e., the accumulating difference in Faraday rotation between neighbouring rays (see Fig.~\ref{fig:differential-precession}). Differential precession in a constant-phase hypersurface is described by the quantity $\chi$, which is a certain combination of NP scalars: $\chi = \beta - \overline{\alpha}$. 

In our discussion of differential precession we pointed out that a local observer, equipped with a locally-straight basis, will interpret differential precession as an additional phase in the geometrical-optics expansion at $O(\omega^{-1})$. The gradient of the \emph{modified} phase leads to a modified tangent vector $K_a$. As the sign of the additional phase follows the handedness of the wave, it appears there is a possible mechanism for left- and right-handed waves to be separated. However, caution is needed: the quantity $\chi$ (which is related to the affine connection in the constant-phase hypersurface) depends on the choice of `slicing'. That is, unlike the optical scalars expansion and shear, $\chi$ is not invariant under $m^a \rightarrow m^a - B k^a$. These issues are investigated further in Ref.~\refcite{Dolan:2018ydp}, where we examine the higher-order (in $\omega$) corrections to the geometrical-optics expansion, and the associated flow of stress-energy. Work is underway to apply these methods to study the `spin-helicity effect' for electromagnetic waves on the Kerr black hole spacetime \cite{Leite:2017zyb}.

\section*{Acknowledgments}
In honour of Prof.~Atsushi Higuchi on the occasion of his 60th birthday. With warm thanks to the organisers of the IV Amazonian Symposium on Physics, held in Bel\'em, Brazil at the Universidade Federal do Par\'a on 18th--22nd September 2017. Special thanks to Luiz Leite, Lu\'is Crispino, Abraham Harte, Antonin Coutant and Jake Shipley for helpful discussions. With additional thanks to the Free University of Sheffield, and Sheffield Central Library, where this work was completed. I acknowledge financial support from the Engineering and Physical Sciences Research Council (EPSRC) under Grant No.~EP/M025802/1, and from the Science and Technology Facilities Council (STFC) under Grant No. ST/L000520/1, and from the project H2020-MSCA-RISE-2017 Grant FunFiCO-777740.

\bibliographystyle{ws-ijmpd}
\bibliography{../go_refs}

\end{document}